\documentclass{ws-ijmpd}
\usepackage[super,compress]{cite}
\usepackage{hyperref}

\begin{document}

\markboth{Llohann D. Speran\c ca}
{An identification of the Dirac operator with the parity operator}

%
\catchline{}{}{}{}{}
%

\title{AN IDENTIFICATION OF THE DIRAC OPERATOR WITH THE PARITY OPERATOR}

\author{LLOHANN D. SPERAN\c CA}

\address{Department of Mathematics,\\
Federal University of Paran\'a,  Curitiba, Paran\'a, Brazil\\
lsperanca@ufpr.br}

\maketitle

\begin{history}
\received{Day Month Year}
\revised{Day Month Year}
\end{history}

\begin{abstract}
We provide a new derivation of the Dirac equation which promptly  generalizes to higher spins. We apply this idea to spin-half Elko dark matter.
\end{abstract}

\def\p{\mbox{\boldmath$\displaystyle\mathbf{p}$}}
\def\r{\mbox{\boldmath$\displaystyle\mathbf{r}$}}
\def\e{\mbox{\boldmath$\displaystyle\mathbf{\epsilon}$}}
\def\L{\mbox{\boldmath$\displaystyle\mathbf{\Lambda}$}}
\def\l{\mbox{\boldmath$\displaystyle\mathbf{\lambda}$}}
\def\th{\mbox{\boldmath$\displaystyle\mathbf{\theta}$}}
\def\k{\mbox{\boldmath$\displaystyle\mathbf{k}$}}
\def\g{\mbox{\boldmath$\displaystyle\mathbf{g}$}}
\def\q{\mbox{\boldmath$\displaystyle\mathbf{q}$}}
\def\bv{\mbox{\boldmath$\displaystyle\mathbf{\varphi}$}}
\def\hp{\mbox{\boldmath$\displaystyle\mathbf{\widehat{\p}}$}}
\def\0{\mbox{\boldmath$\displaystyle\mathbf{0}$}}
\def\O{\mbox{\boldmath$\displaystyle\mathbf{O}$}}
\def\s{\mbox{\boldmath$\displaystyle\mathbf{\sigma}$}}
\def\J{\mbox{\boldmath$\displaystyle\mathbf{J}$}}
\def\K{\mbox{\boldmath$\displaystyle\mathbf{K}$}}
\def\sJ{\mbox{\boldmath$\displaystyle\mathscr{J}$}}
\def\sK{\mbox{\boldmath$\displaystyle\mathscr{K}$}}
\def\mJ{\mbox{\boldmath$\displaystyle\mathcal{J}$}}
\def\iJ{\mbox{\boldmath$\displaystyle\mathit{J}$}}
\def\iK{\mbox{\boldmath$\displaystyle\mathit{K}$}}
\def\mK{\mbox{\boldmath$\displaystyle\mathcal{K}$}}
\def\x{\mbox{\boldmath$\displaystyle\mathbf{x}$}}
\def\y{\mbox{\boldmath$\displaystyle\mathbf{y}$}}
\def\bk{\mbox{\boldmath$\displaystyle\mathbf{\mathcal{K}}$}}
\def\bg{\mbox{\boldmath$\displaystyle\mathbf{\gamma}$}}

\keywords{Dirac operator; Kinematics; ELKO.}

\ccode{PACS numbers: 03.30.+p \and 03.65.Ca}


\section{Introduction}

In his seminal paper\cite{Dirac:1928hu} , Dirac pointed out the incompleteness that was present in previous theories which prevented the introduction of the spin-$\frac{1}{2}$ nature of the electron without further assumptions.  
He solved the problem by introducing Lorentz symmetries to quantum mechanics. In the process, he derived  a first-order differential equation, now known as the \emph{Dirac equation}:
\begin{equation}
(i\gamma^\mu \partial_\mu- m)\psi(x)=0 \label{dirac eq}
\end{equation}
where $\gamma_\mu$ are the complex $4\times 4$ Dirac matrices, $\partial_\mu$ are the derivatives in the space-time directions, $m$ is the mass of the particle and $\psi(x)$ is a \emph{spinor} with four complex entries in the four space-time variables, denoted $x$.
In this paper, we present a new approach to derive the Dirac equation directly from the spin-$\frac{1}{2}$ representation of the Lorentz group without further assumptions and generalize this to higher-spin.
Specifically we show that the parity operator $\mathcal P$ in the spin-$\frac{1}{2}$ representation is given by
\begin{equation}
\mathcal{P}\psi(\vec p\,)=m^{-1}\gamma^\mu p_{\mu}\psi(\vec p\,) \label{equ eq}
\end{equation}
which allows us to derive the Dirac equation from the space-time symmetries alone. This identity provides additional insights on the covariance of the Dirac operator and gives a systematic way of deriving \emph{Dirac-type}  equations on arbitrary $(j,0)\oplus(0,j)$ and $(j,j)$ representations. It also suggests a very simple interpretation of its kinematics principle, namely that kinematics is defined by the conservation of parity which is independent of the frame of reference.

The paper is organized as follows. In Sec.~\ref{1} we derive the parity operator in the $(j,0)\oplus(0,j)$ representation space. Equation (\ref{equ eq}) then follows trivially. In Sec.~\ref{2}, we generalize our derivation to the $(j,j)$ representation and give applications to Elko (see Refs.~[\refcite{Ahluwalia:2004ab,Ahluwalia:2004sz,Ahluwalia:2008xi,Ahluwalia:2009rh}]) in Sec.~\ref{3}.


\section{A Derivation of Dirac Equations from Space-Time Symmetries}\label{1}

Let $\mathfrak{J}_i,\,\mathfrak{K}_{i}$ be rotation and boost generators in the $(\frac{1}{2},0)\oplus(0,\frac{1}{2})$ representation space. 
We write a spinor $\psi(\vec p\,)$ as a function of the momentum in the usual fashion:
\begin{equation}
\label{boost}\psi(\vec p\,)=\exp(i\vec{\mathfrak{K}}\cdot\vec{\varphi})\psi(\vec 0),
\end{equation}
where $\vec{\varphi}=\varphi\hat{p}$ is the rapidity parameter defined as
\begin{equation}
\cosh\varphi=\frac{p^{0}}{m},\hspace{0.5cm}
\sinh\varphi=\frac{p}{m}.
\end{equation}
and the boost matrix is given by
\begin{equation}
\exp(i\vec{\mathfrak{K}}\cdot\vec{\varphi})=
\left(\begin{matrix}
\exp\left(\frac{1}{2}\vec{\sigma}\cdot\vec{\varphi}\right) & O\\
O & \exp\left(-\frac{1}{2}\vec{\sigma}\cdot\vec{\varphi}\right)
\end{matrix}\right)
\end{equation}

The Dirac equation can be derived by simply appealing to the properties of the parity operator  $\mathcal{P}(\vec{p}\,)$. Here we define the operator to be
\begin{equation}
\label{cal P}\mathcal P(\vec p\,)\psi(\vec p\,)=\eta\psi(-\vec p\,)
\end{equation}
where $\eta$ is a $4\times4$ block-off-diagonal matrix
\begin{equation}
\label{eta}\eta=\begin{pmatrix}	O && I \\ 
I && O \end{pmatrix}.
\end{equation}
Using (\ref{boost}) and (\ref{cal P}), we obtain
\begin{eqnarray}
\mathcal P(\vec p\,)\psi(\vec p\,)&=&\eta\psi(-\vec p\,)\\
&=&\eta e^{-2i \vec{\mathfrak{K}}\cdot\vec\varphi}\psi(\vec p\,)\\&=&e^{i\vec{\mathfrak{K}}\cdot\vec\varphi}\eta e^{-i\vec{\mathfrak{K}}\cdot\vec\varphi}\psi(\vec p\,)
\end{eqnarray}
where the last equality follows from the fact that $\eta$ anti-commutes with the generators of boosts. Using the fact that
\begin{equation}\label{id1}\exp(i \vec{\mathfrak{K}}\cdot\vec\varphi)=\cosh\Big(\frac{\varphi}{2}\Big)I+\vec{\sigma}\cdot\hat{p}\sinh\Big(\frac{\varphi}{2}\Big)\end{equation}
the parity operator $\mathcal{P}$ is given by
\begin{eqnarray}\mathcal P(\vec p\,)= m^{-1}\gamma^{\mu}p_{\mu}.
\end{eqnarray}
This operator has eigenvalues $+1$ and $-1$. Taking the spinor $\psi(\vec{p}\,)$ to be eigenspinors of $\mathcal{P}$, we obtain the Dirac equation 
\begin{equation}
(\gamma^{\mu}p_{\mu}\pm mI)\psi(\vec{p}\,).
\end{equation}

\section{Generalization to other Representations}\label{2}
Following the analysis performed in the previous section, the parity operator of the $(j,0)\oplus(0,j)$ representation is
\begin{eqnarray}
\mathcal{P}(\vec{p}\,)&=&\exp(2i\vec{\mathfrak{K}}\cdot\vec{\varphi})\eta\nonumber\\
&=&\frac{1}{m^{2j}}\gamma^{\mu_{1}\cdots\mu_{2j}}p_{\mu_{1}}\cdots p_{\mu_{2j}}.
\end{eqnarray}
where $\vec{\mathfrak{K}}$ is now the boost generator of the $(j,0)\oplus(0,j)$ representation and $\eta$ is a $2(2j+1)\times 2(2j+1)$ block-off-diagonal matrix of the form given by (\ref{eta}). As we will show below, this operator also has eigenvalues $+1$ and $-1$ thus giving us the field equation
for a spinor $\psi(\vec{p}\,)$ of the $(j,0)\oplus(0,j)$ representation space~\cite{Ahluwalia:1991gs,Weinberg:1964cn})
\begin{equation}
\label{DP} 
\left(\frac{1}{m^{2j}}\gamma^{\mu_{1}\cdots\mu_{2j}}p_{\mu_{1}}\cdots p_{\mu_{2j}}\pm I\right)\psi(\vec p\,)=0.
\end{equation}
The operator $\mathcal{P}^{(j)}(\vec{p}\,)$ and the field equation have the following properties
\begin{enumerate}
\item Equation \eqref{DP} has a complete set of solutions: indeed, taking $\vec{p}=\vec{0}$, we get
\begin{equation}\label{uv}u(\vec 0)=\begin{pmatrix}\theta\\\theta\end{pmatrix},\qquad v(\vec 0)=\begin{pmatrix}\theta\\-\theta\end{pmatrix},\end{equation} 
where $[\mathcal{P}(\vec{0}) -I]u(\vec{0})=0$, $[\mathcal{P}(\vec{0}) +I]v(\vec{0})=0$ and that any spinors at rest can be written as a sum of $u(\vec{0})$ and $v(\vec{0})$ in the sense that
\begin{equation}
\begin{pmatrix}\theta\\\lambda
\end{pmatrix}=\frac{1}{2}\left[\begin{pmatrix}\theta+\lambda\\\lambda+\theta\end{pmatrix}+\begin{pmatrix}\theta
-\lambda\\\lambda-\theta\end{pmatrix}\right].
\end{equation}

\item  
The operator $\mathcal{P}(\vec{p}\,)$ has eigenvalues +1 and $-1$. To show this, we consider 
\begin{eqnarray}
[\mathcal{P}(\vec{p}\,)]^{2}=\frac{1}{m^{4j}}\gamma^{\mu_1...\mu_{2j}}\gamma^{\nu_1...\nu_{2j}}p_{\mu_1}...p_{\mu_{2j}}p_{\nu_1}...p_{\nu_{2j}}
\end{eqnarray}
Using the identities from Refs.~[\refcite{Ahluwalia:1991gs,Weinberg:1964cn}], the right-hand side becomes
\begin{equation}
\gamma^{\mu_1...\mu_{2j}}\gamma^{\nu_1...\nu_{2j}}p_{\mu_1}...p_{\mu_{2j}}p_{\nu_1}...p_{\nu_{2j}}
=(p^{\mu}p_{\mu})^{2j}
\end{equation}
which on the mass-shell becomes $m^{4j}$. The eigenvalues of $\mathcal{P}(\vec{p}\,)$ is then just given by its determinant 
\begin{equation}
\det[\mathcal{P}(\vec{p}\,)]=\pm1.
\end{equation}
\end{enumerate} 

Motivated to generalize these framework, we introduce the following definition\\
\\
\noindent\textbf{Definition 3.1}
An operator $\mathcal A(\vec{p}\,)$ in the spinor space is called a \emph{fully kinematic operator} if it satisfies 
\begin{equation}
\mathcal{A}(\Lambda\vec{p})=\mathcal{D}(\Lambda)\mathcal{A}(\vec{p})\mathcal{D}(\Lambda)^{-1}
\label{eqv}
\end{equation}
and the two following conditions
\begin{eqnarray}
&&(1)\,\,\mathcal A(\vec p\,)^2=I \nonumber\\
&&(2)\,\,\{\mathcal A(\vec 0),\vec{\mathfrak K}\}=O
\end{eqnarray}
where $\vec{\mathfrak{K}}$ are the boost generators of the $(j,0)\oplus(0,j)$ representation and $O$ is the zero operator. Here the  condition (2) is required to guarantee that the eigenvalue equation for $\mathcal A$ is indeed a partial differential equation. 

In $(\frac{1}{2},0)\oplus(0,\frac{1}{2})$ representation, all kinematic operators are of the form
\begin{equation}\mathcal A(\vec 0)=\begin{pmatrix}0 &&aI_{2\times 2}\\a^{-1}I_{2\times 2}&&0\end{pmatrix}\end{equation}
where $a$ is a non-zero arbitrary complex number. The requirement that both eigenspaces must be treated in equal footing forces  $a$ to be 1 thus giving us
\begin{equation}
\mathcal{A}(\vec{0})\vert_{a=1}=\eta.
\end{equation}

The  $(j,0)\otimes (0,j)$ representation also admits a fully kinematic operator. Indeed,  $\mathcal A(\vec{0})$, defined by
\begin{equation}\label{par RxL}\mathcal A(\vec 0)(\psi\otimes\varphi)=\varphi\otimes\psi,\end{equation}
and by \eqref{eqv} for other momentum, satisfies all conditions and have the property that $t\mathcal P=\mathcal A t$,
where $t:(j,0)\oplus(0,j)\to (j,0)\otimes(0,j)$ is the (non-linear) tensor product map. In fact, $t\psi$ is a +1 eigenspinor of $\mathcal A$ if $\psi$ is an eigenspinor of $\mathcal P$.

\section{Anti-Linear Kinematic Operators}\label{3}

This framework of kinematic operators recognizes the theory introduced in Ref.~[\refcite{Ahluwalia:2004ab}]. For a dark matter candidate, a structure defined through an anti-linear operator may be of interest, since standard $U(1)$-gauge transformations would not be supported. In this section we restrict ourselves to the spin-$\frac{1}{2}$ case.

However, there is no anti-linear fully kinematic operator. A straightforward calculation shows that the only solutions for $\{\mathcal A(\vec 0),\vec{\mathfrak K}\}=O$ are of the form
\begin{equation}\label{anti1}\mathcal A(\vec 0)=\begin{pmatrix}a\Theta &&0\\0&&b\Theta\end{pmatrix}K\end{equation}
where $K$ is the complex conjugation operator. One can then easily verify that $\mathcal{A}^{2}(\vec p\,)$ cannot be the identity. Indeed, one cannot even guarantee the existence of a complete set of eigenspinors for $\mathcal A(\vec{0})$ for arbitrary $a,b$.

%
%

\subsection{Dark Matter}

Elko spinor fields have been investigated in the last decade\cite{Boehmer:2007dh,Boehmer:2006qq,Boehmer:2010ma,daRocha:2011yr,daRocha:2008we,daRocha:2009gb,daRocha:2013qhu,Dias:2010aa,Liu:2011nb,Basak:2012sn,Lee:2012thesis} and has been further applied to gravity, cosmology, and field theory\cite{Boehmer:2007dh,Boehmer:2006qq,Boehmer:2010ma,daRocha:2011yr,daRocha:2008we,daRocha:2009gb,daRocha:2013qhu,Dias:2010aa,Liu:2011nb,Basak:2012sn,Lee:2012thesis}. We recall that its starting point, in Ref.~[\refcite{Ahluwalia:2004ab}], is to take eigenspinors of the charge conjugation operator instead of the parity operator. Once a choice of   basis of eigenspinors is made, a linear operator is constructed. This resulting linear operator depends on the choice of basis since the charge conjugation operator is anti-linear. We proceed in the same fashion.

We define the charge-conjugation operator as 
\begin{equation}
\label{Cop}\mathcal C=
\begin{pmatrix} O &&i\Theta \\
-i\Theta && O\end{pmatrix}K.
\end{equation} 
If $u,v\in \mathbb{C}^2$ form a basis, then 
\begin{equation}\widetilde u_\pm=\begin{pmatrix}\pm i\Theta u^*\\u\end{pmatrix}
,\qquad \widetilde v_\pm=\begin{pmatrix}\pm i\Theta v^*\\v\end{pmatrix}\end{equation}
is a basis of the spin-$\frac{1}{2}$ space with
\begin{equation}
\mathcal{C}\widetilde u_\pm=\pm \widetilde u,\qquad \mathcal{C}\widetilde v_\pm=\pm \widetilde v.
\label{CC}
\end{equation}
Note that the association $u\mapsto \widetilde u_\pm$ is not linear, making explicit the anti-linearity of $\mathcal{C}$.

Taking $u$ and $v$ as basis vectors, there is a unique linear operator $\mathcal G(u,v)$ such that the identities \eqref{CC} hold, say, the operator which is the diagonal $(1,1,-1,-1)$ in the chosen basis. We recall that $Elko$ theory is constructed by taking $u$ and $v$ as eigenspinors of the helicity operator. In Ref.~[\refcite{Ahluwalia:2010zn}] it is shown that the theory based on $\widetilde{u}_{\pm}$ and $\widetilde{v}_{\pm}$ does not support the full Lorentz group. Here we generalize this fact and prove that no choice of $u$ and $v$ can be done preserving condition \eqref{eqv} for the full Lorentz group.

Taking
\begin{equation} 
u=\begin{pmatrix}a\\b\end{pmatrix}\quad \text{and}\quad v=\begin{pmatrix}c\\d\end{pmatrix}
\end{equation}
we have
\begin{equation}\label{matrix 1}\mathcal G(u,v)=
\begin{pmatrix}
 0 & 0 & \frac{2i {\rm{Im}} \left(b d^*\right)}{a d-b c} & \frac{i \left(c b^*-a d^*\right)}{a d-b c} \\
 0 & 0 & \frac{i \left(d a^*-b c^*\right)}{a d-b c} & \frac{2i{\rm{Im}} \left(a c^*\right)}{a d-b c} \\
 \frac{2i{\rm{Im}} \left(c a^*\right)}{a^* d^*-b^* c^*} & \frac{i \left(c b^*-a d^*\right)}{a^* d^*-b^* c^*} & 0 & 0 \\
 \frac{i \left(d a^*-b c^*\right)}{a^* d^*-b^* c^*} & \frac{2i{\rm{Im}} \left(d b^*\right)}{a^* d^*-b^* c^*} & 0 & 0 \\
\end{pmatrix}
\end{equation}

Now we begin the proof by supposing that we have a theory given by an operator $\mathcal G(\vec p\,)$ satisfying \eqref{eqv}, and that $\mathcal G(u,v)=\mathcal G(\vec p_0)$ for some fixed $\vec p_0$. Using the identity $\Theta\vec{\mathfrak K} \Theta=-\vec{\mathfrak K}^*$  (see Ref.~[\refcite{Ahluwalia:2004ab}]), we get
\[\mathcal G(\vec 0)=e^{-i\vec{\mathfrak K}\cdot\vec\varphi_0} \mathcal G(u,v)e^{i\vec{\mathfrak K}\cdot\vec\varphi_0}=\mathcal G(e^{i\vec{K}\cdot\vec\varphi_0}u, e^{i\vec{K}\cdot\vec\varphi_0}v).\]
Therefore, we can suppose without loss of generality that $\vec p_0=\vec 0$. In this case,  \eqref{eqv} implies that $\mathcal G(u,v)$ must commute with rotations. An application of Schur's Lemma shows that this happens if and only if 
\begin{subequations}
\begin{eqnarray}a\bar d-c\bar b&=&0\label{cond1}\\ \mbox{Im}(a\bar c)&=&\mbox{Im}(b\bar d)\label{cond2}.\end{eqnarray}
\end{subequations}
The first condition, \eqref{cond1}, is equivalent to the existence of $\lambda\in\mathbb{C}$ such that
\[a=\lambda \bar b,\qquad c=\lambda\bar d.\]
In this case we have
\begin{equation}
\det \begin{pmatrix}a&c\\b&d\end{pmatrix}=\det \begin{pmatrix}\lambda\bar b&\lambda\bar d\\b&d\end{pmatrix}=\lambda(\bar bd-\bar db)=2\lambda\mbox{Im}(b\bar d)
\end{equation}
On the other hand, \eqref{cond2} implies that
\begin{equation}
\mbox{Im}(b\bar d)=\mbox{Im}(a\bar c)=\mbox{Im}(|\lambda|^2\bar bd)=-|\lambda|^2\mbox{Im}(b\bar d),
\end{equation}
concluding that $\mbox{Im}(b\bar d)$ or $\lambda$ must be zero. In both cases $u$ and $v$ does not form a basis, as desired.

As a final remark, we observe that the whole construction in Refs.~[\refcite{Ahluwalia:2004ab,Ahluwalia:2004sz}] can be recovered from this kinematic operator viewpoint. In particular, one gets the following decomposition\cite{Ahluwalia:2013uxa}:
\begin{equation}
\gamma^\mu p_\mu=m\mathcal G(\vec p\,)\Xi(\vec p\,)
\end{equation}
 where $\mathcal G(\vec{p}\,)$ is as in Ref.~[\refcite{Ahluwalia:2004ab}] and $\Xi(\vec p\,)$ is the \emph{index-flipping matrix} defined by equation (8) of Ref.~[\refcite{Ahluwalia:2004sz}]. Generally one can chose a set of spin-$j$ spinors $u_{-j},\cdots,u_j,v_{-j},\cdots,v_j$ with norm $\sqrt{2m}$ and consider the operator $\mathcal K(\vec p\,)$ defined as the only operator such that 
\begin{equation}
\mathcal K(\vec p\,)u_\sigma( \vec p\,)=u_\sigma(\vec p\,),\qquad \mathcal K(\vec p\,)v_\sigma(\vec p\,)=
-v_\sigma(\vec p\,)\end{equation}
for all $\sigma$. One notices that, according to our definition of spinors, $\mathcal K(\vec{p}\,)$ always satisfies \eqref{eqv} when $\Lambda$ is the boost in the $\vec p$ direction. Moreover, $\mathcal K(\vec p\,)$ can be defined in the following manner
\begin{equation}
\mathcal K(\vec 0)=\frac{1}{2m}\sum_{\sigma=-j}^{j}\big[u_\sigma(\vec 0)u_{\sigma}^{\dagger}(\vec 0)+v_\sigma(\vec 0)v_{\sigma}^{\dagger}(\vec 0)\big]\widetilde\Xi^{\dagger}(\vec 0)\eta
\end{equation}
where $\Xi(\vec 0)$ is an operator such that 
\begin{eqnarray}
u_\sigma(\vec p\,)^\dagger\widetilde\Xi^{\dagger}(\vec p\,)\eta u_{\sigma'}(\vec p\,)&=&2m\delta_{\sigma\sigma'}\nonumber\\
u_\sigma(\vec p\,)^\dagger\widetilde\Xi^{\dagger}(\vec p\,)\eta v_{\sigma'}(\vec p\,)&=&0\nonumber\\
v_\sigma(\vec p\,)^\dagger\widetilde\Xi^{\dagger}(\vec p\,)\eta v_{\sigma'}(\vec p\,)&=&-2m\delta_{\sigma\sigma'}.\end{eqnarray}
In particular 
\[\widetilde\Xi(\vec p\,)=e^{i\vec{\mathfrak K}\cdot\vec\varphi}\widetilde\Xi(\vec 0)e^{-i\vec{\mathfrak K}\cdot\vec\varphi}.\]
The operator $\widetilde\Xi(\vec 0)$ is uniquely defined and it exists for any set of spinors satisfying \eqref{boost}. A more elaborated definition is required for the case in Ref.~[\refcite{Ahluwalia:2004sz}], although $\Xi(\vec p\,)$ is still well-defined  for $\vec p\neq \vec 0$. 

We now further assume that  $\mathcal K(\vec 0)$ is Hermitian. This is equivalent to ask that the subspaces generated by $u_1,\dots,u_s$ and $v_1,\dots,v_s$ to be Hermitian orthogonal. In this case, keeping the same operator, we can replace the chosen set of spinors by an orthogonal one. Therefore, we get 
\begin{equation}
 \sum_{\sigma=-j}^{j}\big[u_\sigma(\vec 0)u_{\sigma}^{\dagger}(\vec 0)+v_\sigma(\vec 0)
 v_{\sigma}^{\dagger}(\vec 0)\big]=2mI,
\end{equation}
and $\mathcal{K}(\vec{p}\,)$ is given by
\begin{eqnarray}
\mathcal K(\vec p\,)&=&\frac{1}{2m}\sum_{\sigma=-j}^{j}e^{i\vec{\mathfrak K}\cdot\vec\varphi}(e^{i\vec{\mathfrak K}\cdot\vec\varphi})^\dagger (e^{-i\vec{\mathfrak K}\cdot\vec\varphi})^\dagger\widetilde\Xi^{\dagger}(\vec 0)(e^{i\vec{\mathfrak K}\cdot\vec\varphi})^\dagger\eta \nonumber\\
&=&e^{i\vec{\mathfrak K}\cdot\vec\varphi}\widetilde\Xi^{\dagger}(\vec 0)e^{i\vec{\mathfrak K}\cdot\vec\varphi}\eta \nonumber\\
&=&\big[e^{i\vec{\mathfrak K}\cdot\vec\varphi}\widetilde\Xi^{\dagger}(\vec 0)e^{-i\vec{\mathfrak K}\cdot\vec\varphi}\big]e^{i\vec{\mathfrak K}\cdot\vec\varphi}\eta e^{-i\vec{\mathfrak K}\cdot\vec\varphi}\nonumber\\
&=&\Xi(\vec p\,)\mathcal P(\vec p\,)
\end{eqnarray}
where $\Xi(\vec  0)=\widetilde\Xi^\dagger(\vec 0)$. Observing that $\mathcal{K}^{2}(\vec p\,)=1$, we recover the decomposition
\begin{equation}\label{final}
\gamma^\mu p_\mu=m\mathcal K(\vec p\,)\Xi(\vec p\,)
\end{equation}
as desired.

\section*{Conclusions}

In this paper we describe a systematic derivation of the Dirac operator through space-time symmetries and the parity operator. We generalize this procedure, showing how the space-time symmetries produce different partial differential equations out of a special class of operators which we call \emph{fully kinematic operators} (see definition 3.1). 

We apply the same procedure to the (non fully kinematic) charge conjugation operator and conclude that the solutions of the associated linear problem can not be mathematically well-defined at the origin. We further show that the general Parity operator splits in unique way into a generalization of the charge conjugation operator and a new operator $\Xi$.

We believe that the present procedure naturally generalizes to different contexts including different symmetry groups. For instance, we observe that it certainly generalize to Minkowski $(n+1)$-space with $O(n,1)$. Since Fourier transformation interchanges eigenvectors of the associated linear problem with solution to its derived differential equations, we expect that this method can provide solutions and interesting properties to many known and unknown equations. 

\section*{Acknowledgements}

The operator given in \eqref{equ eq} was implicitly identified as the parity operator by D.~V.~Ahluwalia in his derivation of the Dirac equation~\cite{Ahluwalia:1991gs}. The author is thankful to him for the reference and further discussions, to Cheng-Yang Lee for a complete revision of the text and to the anonymous referee for further suggestions and corrections.

\label{Bibliography}
\bibliographystyle{ws-ijmpd}
\bibliography{Bibliography}

\begin{thebibliography}{10}

\bibitem{Dirac:1928hu}
P.~A.~M. Dirac, {\em Proc. Roy. Soc. Lond.} {\bf A117}  (1928) 610.

\bibitem{Ahluwalia:2004ab}
D.~V. Ahluwalia and D.~Grumiller, {\em JCAP} {\bf 0507}  (2005)   012,
  \href{http://arxiv.org/abs/hep-th/0412080}{{\ttfamily arXiv:hep-th/0412080}}.

\bibitem{Ahluwalia:2004sz}
D.~V. Ahluwalia and D.~Grumiller, {\em Phys. Rev.} {\bf D72}  (2005)   067701,
  \href{http://arxiv.org/abs/hep-th/0410192}{{\ttfamily arXiv:hep-th/0410192}}.

\bibitem{Ahluwalia:2008xi}
D.~V. Ahluwalia, C.-Y. Lee and D.~Schritt, {\em Phys.Lett.} {\bf B687}  (2010)
  248, \href{http://arxiv.org/abs/0804.1854}{{\ttfamily arXiv:0804.1854
  [hep-th]}}.

\bibitem{Ahluwalia:2009rh}
D.~V. Ahluwalia, C.-Y. Lee and D.~Schritt, {\em Phys.Rev.} {\bf D83}  (2011)
  065017, \href{http://arxiv.org/abs/0911.2947}{{\ttfamily arXiv:0911.2947
  [hep-ph]}}.

\bibitem{Ahluwalia:1991gs}
D.~V. Ahluwalia, {\em Doctoral thesis, Texas A\&M University}   (1991)   241 p.

\bibitem{Weinberg:1964cn}
S.~Weinberg, {\em Phys. Rev.} {\bf 133}  (1964) B1318.

\bibitem{Boehmer:2007dh}
C.~G. Boehmer, {\em Annalen Phys.} {\bf 16}  (2007) 325,
  \href{http://arxiv.org/abs/gr-qc/0701087}{{\ttfamily arXiv:gr-qc/0701087
  [gr-qc]}}.

\bibitem{Boehmer:2006qq}
C.~G. Boehmer, {\em Annalen Phys.} {\bf 16}  (2007) 38,
  \href{http://arxiv.org/abs/gr-qc/0607088}{{\ttfamily arXiv:gr-qc/0607088
  [gr-qc]}}.

\bibitem{Boehmer:2010ma}
C.~G. Boehmer, J.~Burnett, D.~F. Mota and D.~J. Shaw, {\em JHEP} {\bf 1007}
  (2010)   053, \href{http://arxiv.org/abs/1003.3858}{{\ttfamily
  arXiv:1003.3858 [hep-th]}}.

\bibitem{daRocha:2011yr}
R.~da~Rocha, A.~E. Bernardini and J.~Hoff~da Silva, {\em JHEP} {\bf 1104}
  (2011)   110, \href{http://arxiv.org/abs/1103.4759}{{\ttfamily
  arXiv:1103.4759 [hep-th]}}.

\bibitem{daRocha:2008we}
R.~da~Rocha and J.~M. Hoff~da Silva, {\em Adv.Appl.Clifford Algebras} {\bf 20}
  (2010) 847, \href{http://arxiv.org/abs/0811.2717}{{\ttfamily arXiv:0811.2717
  [math-ph]}}.

\bibitem{daRocha:2009gb}
R.~da~Rocha and J.~Hoff~da Silva, {\em Int.J.Geom.Meth.Mod.Phys.} {\bf 6}
  (2009) 461, \href{http://arxiv.org/abs/0901.0883}{{\ttfamily arXiv:0901.0883
  [math-ph]}}.

\bibitem{daRocha:2013qhu}
R.~da~Rocha, L.~Fabbri, J.~Hoff~da Silva, R.~Cavalcanti and J.~Silva-Neto, {\em
  J. Math. Phys.} {\bf 54}  (2013)   102505,
  \href{http://arxiv.org/abs/1302.2262}{{\ttfamily arXiv:1302.2262 [gr-qc]}}.

\bibitem{Dias:2010aa}
M.~Dias, F.~de~Campos and J.~Hoff~da Silva, {\em Phys.Lett.} {\bf B706}  (2012)
  352, \href{http://arxiv.org/abs/1012.4642}{{\ttfamily arXiv:1012.4642
  [hep-ph]}}.

\bibitem{Liu:2011nb}
Y.-X. Liu, X.-N. Zhou, K.~Yang and F.-W. Chen, {\em Phys.Rev.} {\bf D86}
  (2012)   064012, \href{http://arxiv.org/abs/1107.2506}{{\ttfamily
  arXiv:1107.2506 [hep-th]}}.

\bibitem{Basak:2012sn}
A.~Basak, J.~R. Bhatt, S.~Shankaranarayanan and K.~Prasantha~Varma, {\em JCAP}
  {\bf 1304}  (2013)   025, \href{http://arxiv.org/abs/1212.3445}{{\ttfamily
  arXiv:1212.3445 [astro-ph.CO]}}.

\bibitem{Lee:2012thesis}
C.-Y. Lee, {\em Doctoral thesis, University of Canterbury}   (2012)   198 p.

\bibitem{Ahluwalia:2010zn}
D.~V. Ahluwalia and S.~P. Horvath, {\em JHEP} {\bf 1011}  (2010)   078,
  \href{http://arxiv.org/abs/1008.0436}{{\ttfamily arXiv:1008.0436 [hep-ph]}}.

\bibitem{Ahluwalia:2013uxa}
D.~V. Ahluwalia  (2013) \href{http://arxiv.org/abs/1305.7509}{{\ttfamily
  arXiv:1305.7509 [hep-th]}}.

\end{thebibliography}

\end{document}